\documentclass[10pt,aps,prl,twocolumn,groupedaddress,superscriptaddress,showkeys]{revtex4-2}
\usepackage{amsmath}
\usepackage{graphicx}
\DeclareGraphicsExtensions{.pdf,.eps,.png,.jpg,.mps}
\DeclareUnicodeCharacter{2061}{}
\usepackage{epstopdf}
\usepackage{color}
\usepackage{float}
\usepackage{soul}
\usepackage{amssymb}
\usepackage[normalem]{ulem}
\usepackage{todonotes}
\usepackage{booktabs}

\begin{document}
\title{Graphene phase modulators operating in the transparency regime}
\author{H. F. Y. Watson}
\author{A. Ruocco}
\author{M. Tiberi}
\author{J. E. Muench}
\author{O. Balci}
\author{S. M. Shinde}
\author{S. Mignuzzi}
\affiliation{Cambridge Graphene Centre, University of Cambridge, 9 JJ Thomson Avenue, CB3 0FA, Cambridge, UK}
\author{M. Pantouvaki}
\author{D. Van Thourhout}
\affiliation{IMEC, Kapeldreef 75, B-3001 Leuven, Belgium}
\author{R. Sordan}
\affiliation{Politecnico di Milano, Polo di Como, Via Anzani 42, 22100 Como, Italy}
\author{A. Tomadin}
\affiliation{Dipartimento di Fisica, Universit\`{a} di Pisa, Largo Bruno Pontecorvo 3, 56127 Pisa, Italy}
\author{M. Romagnoli}
\affiliation{Consorzio Nazionale per le Telecomunicazioni, 56124 Pisa, Italy}
\author{A. C. Ferrari}
\affiliation{Cambridge Graphene Centre, University of Cambridge, 9 JJ Thomson Avenue, CB3 0FA, Cambridge, UK}
\thispagestyle{empty}
\pagestyle{empty}
\begin{abstract}
Next-generation data networks need to support Tb/s rates. In-phase and quadrature (IQ) modulation combine phase and intensity information to increase the density of encoded data, reduce overall power consumption by minimising the number of channels, and increase noise tolerance. To reduce errors when decoding the received signal, intersymbol interference must be minimised. This is achieved with pure phase modulation, where the phase of the optical signal is controlled without changing its intensity. Phase modulators are characterised by the voltage required to achieve a $\pi$ phase shift V$_{\pi}$, the device length L, and their product V$_{\pi}$L. To reduce power consumption, IQ modulators are needed with$<$1V drive voltages and compact (sub-cm) dimensions, which translate in V$_\pi$L$<$1Vcm. Si and LiNbO$_3$ (LN) IQ modulators do not currently meet these requirements, because V$_{\pi}$L$>$1Vcm. Here, we report a double single-layer graphene (SLG) Mach-Zehnder modulator (MZM) with pure phase modulation in the transparent regime, where optical losses are minimised and remain constant with increasing voltage. Our device has $V_{\pi}L\sim$0.3Vcm, matching state-of-the-art SLG-based MZMs and plasmonic LN MZMs, but with pure phase modulation and low insertion loss ($\sim$5dB), essential for IQ modulation. Our $V_\pi L$ is$\sim$5 times lower than the lowest thin-film LN MZMs, and$\sim$3 times lower than the lowest Si MZMs. This enables devices with complementary metal-oxide semiconductor compatible V$_\pi$L ($<$1Vcm) and smaller footprint than LN or Si MZMs, improving circuit density and reducing power consumption by one order of magnitude.
\end{abstract}
\maketitle
The global internet traffic was expected to triple between 2019-2024 with the advent of 5G and the internet of everything\cite{Cisco2019}. Lockdowns in response to COVID-19 shifted the distribution of data traffic across the network\cite{Smaragdakis2023}, with an additional$\sim$20-200\% rise\cite{Feldmann2021}, due to remote working\cite{Feldmann2021, Marcus2023}, and increased use of home entertainment services\cite{Ericcson2020}. This vast amount of data relies on a backbone of high-density data network infrastructures, with 2018 standards of 400Gb/s\cite{IEEEComputerSociety2018} to be extended$>$1Tb/s by 2025\cite{EthernetAlliance2023}. To go$>$1Tb/s it is preferable to increase data rates in a single channel\cite{Wei2015,Sharif2015}, rather than the number of channels. By minimising the number of channels, the power consumption and system complexity is reduced, because less electrical drivers and active optical components are needed\cite{Wei2015,Sharif2015}. The bandwidth (BW) of a single channel that uses binary modulation formats is limited by that of the electrical interfaces used to drive the active optical components\cite{Wei2015,Sharif2015}. These struggle to exceed 2023 standards\cite{EthernetAlliance2023} because losses increase with increased frequencies\cite{Miller2000}. Consequently, for data rates$>$100Gb/s, binary modulation formats\cite{KumarShiva2014FOC} have been replaced by 4-level pulse-amplitude modulation (PAM)\cite{IEEEComputerSociety2018}. PAM uses 4 amplitude levels of the transmitted optical signal, to represent 4 symbols that correspond to 2 bits of information\cite{Wei2015}. Other multi-level modulation schemes, such as quadrature amplitude modulation (QAM)\cite{KumarShiva2014FOC}, encode information in both phase and amplitude\cite{KumarShiva2014FOC}. Transmission systems that use only amplitude modulation (AM) are known as direct detection systems\cite{KumarShiva2014FOC}. Those that use both phase modulation (PM) and AM are known as coherent, because the phase difference between two or more signals remains constant over time\cite{KumarShiva2014FOC}. Coherent systems have a higher noise tolerance than direct detection ones, because the signal degradation from fibre dispersion can be compensated by the received signal phase\cite{KumarShiva2014FOC}.
\begin{table*}[t]
    \centering
    \begin{tabular*}{\linewidth}{@{\extracolsep{\fill}}ccccccccc}
    \toprule % Top horizontal line
    Ref. & Material & Type & IL [dB] & ER [dB] & V$_\pi$L [Vcm] & L [cm] & Modulation speed & V$_\pi$IL [VdB] \\ % Column names row
    \midrule % In-table horizontal line
    \cite{Li2018} & Si & MZM & 5.4 & 3.6 & 1.4 & 0.2 & 55 GHz & 38 \\
    \cite{Sun2019} & Si & Ring modulator & 3 & 9.8 & 0.52 & 0.2 & 50 GHz & 8 \\
    \cite{Harris2014} & Si & Thermo-optic PM & 0.23 & - & 0.027 & 0.006 & 130 kHz & 1 \\
    \cite{Chen2011} & III-V/Si MOS & MZM depletion mode & - & 11 & 0.24 & 0.05 & 27 GHz & - \\
    \cite{Han2017} & III-V/Si MOS & MZM accumulation mode & 1 & 12 & 0.047 & 0.047 & 100 MHz & 1 \\
    \cite{Hiraki2018} & III-V/Si MOS & MZM depletion mode & - & 4.4 & 0.3 & 0.03 & 18 GHz & 70 \\
    \cite{Wang2018a} & Thin-film LN & MZM & 0.5 & 30 & 1.4 & 2 & $>$ 45 GHz & 0.4 \\
    \cite{Weigel2018} & Thin-film LN & MZM & 7.6 & 20 & 6.7 & 0.5 & 106 GHz & 102 \\
    \cite{He2019} & Thin-film LN/Si & MZM & 2.5 & 40 & 2.2 & 0.3 & $>$ 70 GHz & 18.5 \\
    \cite{Li2023} & Thin-film LN/Si & MZM & 15 & 19 & 0.8 & 0.3 & $>$ 40 GHz & 19.5 \\
    \cite{Thomaschewski2022} & Thin-film LN & Plasmonic MZM & 19.5 & 2.5 & 0.23 & 0.0015 & $>$ 10 GHz & 2,990 \\
    \cite{Giambra2019} & DSLG & EAM & 20 & 3 & - & 0.01 & 29 GHz & - \\
    \cite{Phare2015} & DSLG & Ring modulator & - & 15 & - & 0.003 & 30 GHz & - \\
    \cite{Agarwal2021} & DSLG (flakes) & EAM & 4 & 5 & - & 0.006 & 39 GHz & - \\
    \cite{Sorianello2018} & SLG/Si & MZM & 10 & 35 & 0.28 & 0.03 & 5 GHz & 62 \\
    This work & DSLG & MZM & 5.6 & 25 & 0.3 & 0.0075 & 24 GHz & 3 \\
    \bottomrule % Bottom horizontal line
    \end{tabular*}
    \caption{Modulators based on Si, III-V (InGaAsP), LN and graphene for IQ modulators design.}
    \label{tab1:Modulators}
\end{table*}

Information is transmitted by electro-optic (EO) modulators that convert an electrical signal into an optical one\cite{ReedGrahamT2004Sp:a}. This can be encoded into the intensity of the transmitted signal, known as AM, or electro-absorption modulation\cite{ReedGrahamT2004Sp:a}, and into the phase, known as PM or electro-refractive modulation\cite{ReedGrahamT2004Sp:a}. In-phase and quadrature (IQ) modulators are interferometric devices that use pure PM, with no change of amplitude, to generate the different QAM symbols\cite{KumarShiva2014FOC}. No direct AM is required to generate QAM symbols, because the interferometer converts a phase difference into a change in amplitude\cite{KumarShiva2014FOC}. To reduce intersymbol interference, therefore errors at the receiver\cite{KumarShiva2014FOC}, the symbol noise should be minimised, and symbols should be evenly spaced in the in-plane and quadrature axes\cite{KumarShiva2014FOC}. Thus, any unwanted AM will increase symbol noise, and any non-linear PM will result in irregular symbol spacing\cite{KumarShiva2014FOC}.

An important parameter for comparing phase modulators is the product of the voltage required to achieve a $\pi$ phase shift, $V_\pi$, and the device length, $L$\cite{ReedGrahamT2008Sp:t}. The additional optical loss resulting from inserting the device in the transmission line is the insertion loss IL=$\alpha L$, where $\alpha$ is the absorption coefficient per unit length\cite{Reed2010}. In order to reduce overall power consumption, we need to minimise $V_\pi L$ and IL\cite{Reed2010}, because a lower $V_\pi$L reduces the device area and capacitance, hence reducing the dynamic energy consumption E=CV$^2$=CV$_{pp}^2$/4\cite{Ye2017}, i.e. the energy charged and discharged in a capacitor by an AC voltage with peak-to-peak voltage V$_{pp}$. IL contributes to optical power loss and signal degradation. The PM figure of merit (FOM$_\text{PM}$) is defined as the product of $V_\pi$ and IL (FOM$_\text{PM}=V_\pi\alpha L$)\cite{Tu2011}, whereby better phase modulators have a smaller FOM$_\text{PM}$. The modulator BW is critical for Tb/s data transmission, in order to maximise the data rates that a single channel can support\cite{Reed2010}, which is $T = BWlog_2(1+\frac{S}{N})$\cite{Shannon1948}, where BW is in Hz and the signal to noise ratio, S/N, is the ratio of the received power to the noise power. E.g., a data rate of 100Gb/s in a single lane with S/N$\geq$20, which is the goal set by the 2023 Ethernet Alliance roadmap\cite{EthernetAlliance2023}, requires BW$\geq$23GHz.

Table \ref{tab1:Modulators} shows silicon photonics (SiP)\cite{Li2018, Sun2019, Harris2014}, III-V (InGaAsP)\cite{Chen2011,Han2017,Hiraki2018}, LiNbO$_3$ (LN)\cite{Wang2018a,Weigel2018,He2019,Li2023,Thomaschewski2022} and graphene-based Electro-absorption modulators (EAMs)\cite{Giambra2019,Phare2015,Agarwal2021} and PM\cite{Sorianello2018}. SiP offers a cost-effective solution for integrating electronic and photonic components in the same circuit by using existing complementary metal oxide semiconductor (CMOS) technology\cite{Thomson2016}. Pure PM is difficult to achieve with Si modulators based on the plasma dispersion effect\cite{Xu2005,Liao2005,Thomson2012,Li2018,Sepehrian2019,Sun2019} because, due to the Kramers-Kronig relations\cite{Soref1987}, any change in carrier concentration results in changes in both absorption and phase. Even if pure phase modulators in Si were to be achieved, these devices would rely on doped Si waveguides (WGs), requiring an increased optical power to overcome the additional optical losses introduced by dopants\cite{Soref1987}, when compared to undoped Si WGs. Other modulation mechanisms in Si can be used, such as the thermo-optic effect\cite{Cocorullo1992,Harris2014}, changing the Si optical properties via electrically induced temperature changes. The thermal time constant of Si is$\sim1$ms\cite{Hidalga2000} at RT, limiting operating speeds to the kHz range\cite{Harris2014}. Non-linear effects, such as the Kerr effect\cite{ReedGrahamT2004Sp:a}, produce a change in the refractive index proportional to the product of the nonlinear refractive index and the intensity of the propagating light\cite{ReedGrahamT2004Sp:a}. However, at telecom wavelengths (1.3, 1.5$\mu$m) this is$\sim$3 orders of magnitude weaker than the plasma dispersion effect\cite{Soref1987}. Thus, new materials with higher nonlinear refractive index are needed.

Hybrid approaches that incorporate III-V compounds\cite{Chen2011,Hiraki2017,Han2017,Hiraki2018} with doped Si WGs reduce $V_\pi L$ by utilising other effects, such as band-filling\cite{Numai2015}, which results in reduced absorption due to occupied energy states\cite{Bennett1990}. III-V/Si metal-oxide-semiconductor (MOS) Mach-Zehnder modulators (MZM) operating in accumulation mode\cite{Hsu2022}, which rely on the change in accumulated charge carriers within the MOS capacitor by applying a gate voltage, have the lowest $V_\pi L\sim$0.047 V$\cdot$cm to date\cite{Han2017}, with IL$\sim$1dB\cite{Han2017}, but are BW limited to$\sim$100MHz\cite{Han2017}. III-V/Si MOS MZMs struggle to maintain $V_\pi L=0.047$Vcm with a higher BW, because of the high ($\sim$3$k\Omega\mu$m\cite{Han2017}) contact resistance\cite{Han2017} to the Si electrode in the MOS configuration, with more moderate values of $V_\pi L$=0.24-0.3V$\cdot$cm\cite{Chen2011,Hiraki2018} for III-V MZMs operating in depletion mode with BW up to$\sim$27GHz\cite{Chen2011}. III-V based MZMs offer a lower $V_\pi L$ compared to Si MZMs, but at the cost of more complex fabrication, with expensive III-V processing\cite{Roelkens2007}. Cost-effectiveness is determined by the cost per unit Watt used to manufacture III-V devices, which is $\$$40/W to $\$$100/W at 2018 prices\cite{Horowitz2018,IIIVcosts2022}. This is at least two orders of magnitude higher than Si manufacturing\cite{Horowitz2018,IIIVcosts2022}.

Integrating LiNbO$_3$ (LN) on undoped Si WGs enables pure PM, exploiting the Pockel's effect\cite{KumarShiva2014FOC}, producing a change in refractive index proportional to the electric field. Modulators based on sub-$\mu$m thin-film LN\cite{Wang2018a,Weigel2018,Wang2018,He2019,Li2023} have IL$<$1dB\cite{Wang2018a} and BW$>$100GHz\cite{Wang2018a,Weigel2018}. Thin-film LN MZMs were reported with $V_\pi L\sim$1.4V$\cdot$cm\cite{Wang2018a}, a factor of 2 larger than state-of-the-art Si plasma-dispersion MZMs\cite{Liu2020}. However, this $V_\pi L$ means that cm long devices are needed to reduce $V_\pi$ to CMOS compatible levels$<$1V\cite{HuChenming2010Msdf}. Thin-film LN MZMs with $V_\pi L\sim$0.8Vcm\cite{Li2023} were demonstrated in the visible range, but with IL$\sim$15dB\cite{Li2023}. Plasmonic LN modulators show $V_\pi L\sim$0.23Vcm\cite{Thomaschewski2022}, but with IL$\sim$19.5dB\cite{Thomaschewski2022}. Modulators with lower $V_\pi L$ and IL are essential to increase the density of SiP integrated circuits, thus reducing power consumption by minimising electrical interconnects. The interconnect losses are frequency ($f$) dependent ($\propto\sqrt{f}$\cite{HorowitzP1980Taoe}) due to increased resistance caused by the skin effect\cite{HorowitzP1980Taoe}, where more of the current flows at the surface as $f$ increases\cite{Popovic1999}. Therefore, for electrical interfaces driving Tb/s data rates, the power consumption of interconnects becomes the limiting factor\cite{HorowitzP1980Taoe,Miller2000,Meindl2002}.

Graphene is ideal for opto-electronics\cite{Bonaccorso2010,Koppens2014,Ferrari2015,Romagnoli2018} due to its high carrier mobility ($\mu>$50,000cm$^2$/Vs at room temperature, $RT$\cite{Purdie2018,DeFazio2019}), electrically tunable optical conductivity\cite{Wang2008,Li2008}, and wavelength independent absorption in the visible (500nm) to mid-infrared (10$\mu$m)\cite{Nair2008,Li2008}. The gapless band structure with massless Dirac Fermions in single-layer graphene (SLG) enables the optical conductivity to be electrostatically controlled\cite{Wang2008,Li2008}, and absorption to be suppressed\cite{Pisana2007}. Double SLG (DSLG) phase modulators can reach a theoretical $V_\pi L\sim$0.1V$\cdot$cm\cite{Sorianello2015}, which enables mm long devices with driving voltages$<$1V. When absorption is suppressed, the optical loss can be reduced by orders of magnitude from$>$1000 dB/cm\cite{Sorianello2015} to$<$10dB/cm\cite{Sorianello2015}. The combination of mm lengths and$<$10dB/cm optical losses, leads to IL$<$1dB, therefore minimising power consumption. SLG can be produced at wafer scale\cite{Bonaccorso2012,Ferrari2015,Backes2020}. Chemical vapour deposition (CVD) can be used to grow polycrystalline films up to 30"\cite{Bae2010} or single crystals\cite{Miseikis2017}. The latter allows one to fabricate devices at predetermined locations\cite{Giambra2019,Muench2019}. SLG films can be integrated in the CMOS back-end-of-line for wafer scale processing after fabrication of the integrated circuits\cite{Wu2023}. This can reduce cost and complexity of fabrication, by removing the need for doped Si WGs in DSLG designs\cite{Liu2012,Mohsin2015,Phare2015,Dalir2016,Giambra2019}. EAMs\cite{Liu2011a,Liu2012,Dalir2016,Giambra2019, Agarwal2021} and electro-refractive modulators\cite{Mohsin2015,Phare2015,Sorianello2016,Sorianello2018} (ERMs) based on one or more SLG have been reported, with $V_\pi L\sim$0.28$V\cdot$cm\cite{Sorianello2018} and data transmission rates$\sim$50Gb/s\cite{Giambra2019}. However, pure PM has not been reported yet, to the best of our knowledge.

The SLG conductivity $\sigma(\omega)$, derived from the Kubo formula\cite{Kubo1957}, is a function of the angular frequency of light ($\omega$), SLG transport relaxation time ($\tau$), SLG Fermi level ($E_\text{F}$), and temperature $T$\cite{DasSarma2011,Kotov2012,Chang2014}:
\begin{equation}
\label{optical_conductivity}
\begin{split}
\sigma(\omega)=\frac{\sigma_0}{2}\left[\text{tanh}\left(\frac{\hbar\omega + 2E_\text{F}}{4k_BT}\right) + \text{tanh}\left(\frac{\hbar\omega - 2E_\text{F}}{4k_BT}\right)\right]\\
- i\frac{\sigma_0}{2\pi}\text{log}\left[\frac{(\hbar\omega + 2E_\text{F})^2}{(\hbar\omega - 2E_\text{F})^2 + (2k_BT)^2}\right]\\
+ i\frac{4\sigma_0}{\pi}\frac{E_\text{F}}{\hbar\omega + i\hbar/\tau}
\end{split}
\end{equation}
where $\sigma_0=e^2/4\hbar$ is the $f$-independent, or universal conductivity of SLG\cite{Kuzmenko2008,Nair2008}, $\hbar$ is the reduced Planck's constant, and $k_B$ is the Boltzmann constant. The first two terms represent interband transitions\cite{Hanson2008, Falkovsky2007}. The third represents intraband transitions\cite{Hanson2008, Falkovsky2007}, and it is a function of $\sigma_0$, $E_\text{F}$, $\omega$ and $\tau$. The intraband contribution to $\sigma(\omega)$ can be simplified to express the DC conductivity of SLG ($\sigma_\text{d.c.}$) when $\omega\to$0\cite{Romagnoli2018}. $\tau$ can then be related to $\mu$ by using $\sigma_\text{d.c.}=ne\mu$\cite{KittelC1996Itss}, where $n$ is the carrier concentration given by $E_\text{F}=\hbar v_F\sqrt{n\pi}$\cite{Das2008,DasSarma2011,Kotov2012}. We thus arrive at $\mu\sim e\tau v_F^2/E_\text{F}$\cite{Romagnoli2018} for $E_\text{F}\gg k_BT$, where $v_F\simeq9.5\times10^5\text{ ms}^{-1}$ is the Fermi velocity\cite{Hanson2008,DasSarma2011,Kotov2012}. Eq.\ref{optical_conductivity} implies that $\sigma(\omega)$ of each SLG depends on $E_\text{F}$, and the energy of the incident light ($E_\text{in}=hc/\lambda$). Absorption in undoped SLG is dominated by interband transitions and is suppressed when $2E_\text{F}>hc/\lambda$, due to Pauli blocking\cite{Pisana2007}. For $\lambda$=1.55$\mu$m, or $E_\text{in}$=0.8eV, Pauli blocking occurs for $E_\text{F}>$0.4eV.

For Pauli blocking, SLG enters the transparency regime, whereby interband transitions are suppressed and only intraband transitions occur\cite{Pisana2007}. Intraband transitions dominate for low energy photons ($\omega<$2000cm$^{-1}$\cite{Li2008}) and for $2E_\text{F}>hc/\lambda$. Intraband transitions are dependent on $\tau$ because they depend on scattering centres (e.g. defects) for conservation of momentum\cite{Mak2008}. Therefore, absorption by intraband transitions increases for shorter $\tau$, which is related to mobility $\mu=\frac{e\tau v_F^2}{E_F}$\cite{Romagnoli2018}.

Operating beyond Pauli blocking is essential for pure PM, because in this regime SLG absorption is minimised and constant with respect to gate voltage, thus reducing the overall IL. A DSLG modulator can work as EAM or ERM depending on bias\cite{Gosciniak2013a}. For EAMs, the onset of Pauli blocking results in the largest change in absorption\cite{Liu2011a}, hence the bias should be set at the onset of Pauli blocking. For ERMs, the bias is set beyond the Pauli blocking condition, where the change in refractive index is quasi-linear\cite{Sorianello2016} and absorption is minimised\cite{Sorianello2015}.

Here, we report DSLG-based MZMs on undoped Si WGs operating beyond Pauli blocking with $V_\pi L\sim$0.3V$\cdot$cm and pure PM. These work at 16V without dielectric breakdown, enabling access to the transparent regime. This work represents a key step in the development of graphene-based coherent integrated transmitters for communication systems.
\begin{figure*}
\centerline{\includegraphics[width=\textwidth]{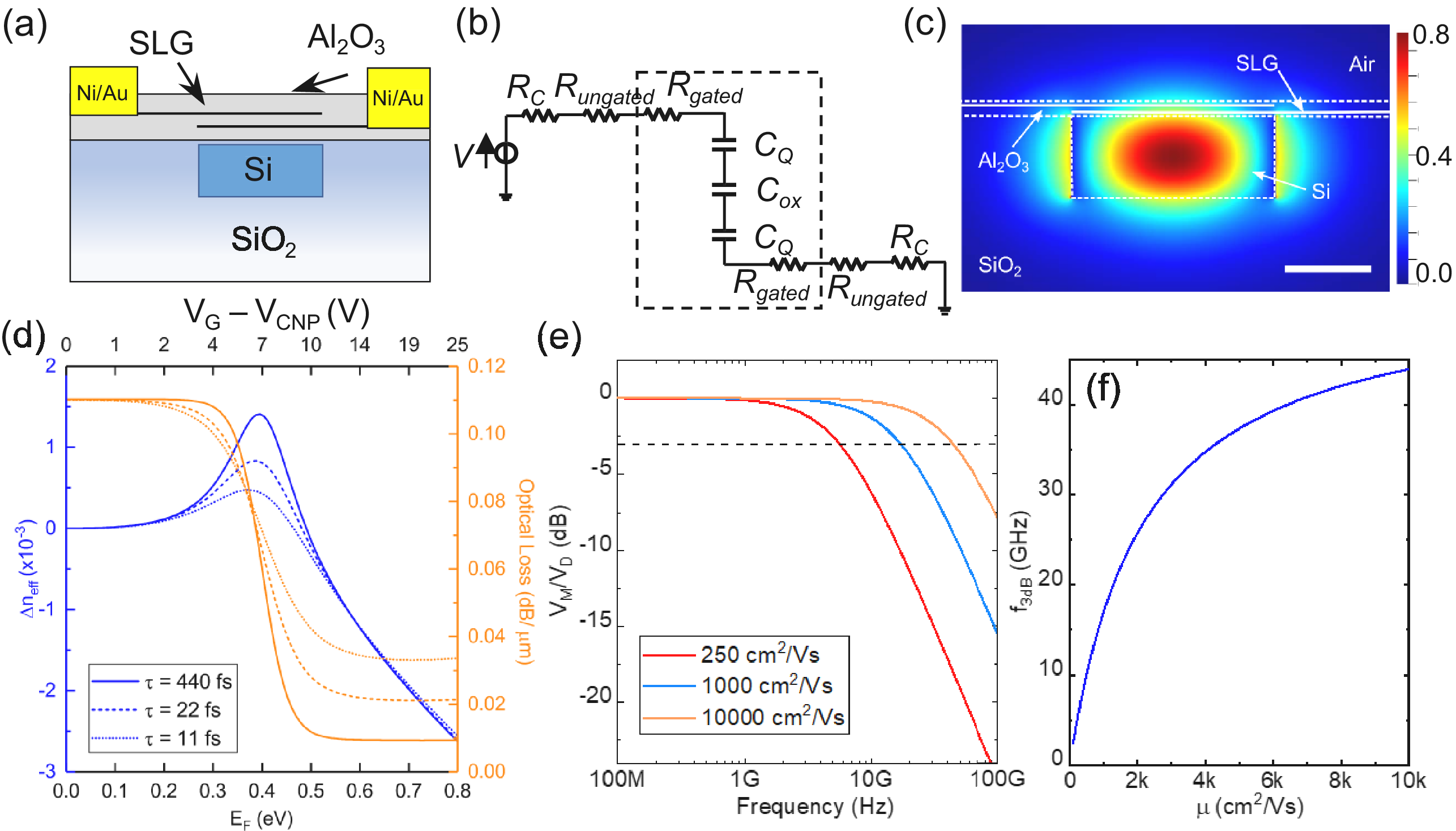}}
\caption{(a) DSLG modulator scheme, with 20nm Al$_2$O$_3$ between bottom and top SLG. (b) Equivalent circuit, used to calculate $Z_\text{T}(\omega)$, $Z_\text{C}(\omega)$ and 3dB cut-off BW $f_{3dB}$. The components contained within the dashed line contribute to the impedance of the overlapping SLG section. (c) Simulated $E_x$ of fundamental TE mode confined within a 550x220nm$^2$ Si WG at 1.55$\mu m$. Colour scale indicates the $E_x$ amplitude, scale bar 200nm. (d) Simulated change in n$_\text{eff}$ (blue) and optical loss (orange) of confined mode due to different $V$-$V_\text{CNP}$ across the DSLG capacitor. Simulations are performed at 1.55$\mu$m ($\sim$0.8eV). SLGs are separated by 20nm Al$_2$O$_3$. The overlapping SLG region is 550nm, the ungated SLG region is 1$\mu$m ($E_\text{F}$=0.2eV). (e) Simulated $f_\text{3\text{dB}}$ as a function of $\mu$ for a 50$\mu m$ DSLG modulator with ungated sections of each SLG$\sim$1$\mu$m ($E_\text{F}\sim$0.23eV), gated sections of each SLG$\sim$450nm ($E_\text{F}\sim$0.4eV). $R_C\sim$1000$\Omega\cdot\mu$m, 20nm Al$_2$O$_3$ with $\epsilon_\text{r}$=8, $C_\text{eq}$ calculated with an additional carrier concentration$\sim10^{10}\text{cm}^{-2}$ from defects and$\sim10^{11}\text{cm}^{-2}$ from charged impurities. $\mu$ calculated at 0.4eV. (f) Simulated $f$ response of DSLG modulator for different $\mu$ for the same modulator specification as (e).}
\label{fig:Figure 1}
\end{figure*}
\section{Results and Discussion}
The design of our DSLG phase modulator is in Fig.\ref{fig:Figure 1}a. It consists of two SLG encapsulated by Al$_2$O$_3$, overlapping in the region above the Si WG. 10nm Al$_2$O$_3$ encapsulates both SLGs to protect them during subsequent processing steps, minimise contamination, and preserve $\mu$. The bottom encapsulation is used to maintain symmetry between the two SLGs, so that both are in the same environment. Each SLG is contacted by a metal placed on either side of the WG. The two SLG layers form a capacitor (equivalent electrical circuit in Fig.\ref{fig:Figure 1}b.), where an applied voltage across the contacts creates a perpendicular electric field which modulates the carrier density, thus $\sigma(\omega)$ of each SLG. This, in-turn, modulates the complex effective refractive index, $n_{\text{eff}}$\cite{Falkovsky2007, Hanson2008}, leading to a change in phase and absorption of light along the propagation direction\cite{ReedGrahamT2004Sp:a}. We simulate the optical performance of our DSLG modulators using the Finite-Difference Eigenmode (FDE) solver in Lumerical\cite{Zhu2002}. This uses the expansion method to calculate the eigenmodes and eigenvalues of Maxwell's equations in the $f$ domain\cite{Bienstman2001}. Each solution, or mode, has its own electromagnetic field profile and $n_{\text{eff}}$\cite{ReedGrahamT2004Sp:a}. The real component is the effective refractive index, $n_\text{eff}$, related to the phase, $\phi$, of the light along $L$ by $\phi=k_0n_\text{eff}L$\cite{ReedGrahamT2004Sp:a}, where $k_0$ ($2\pi/\lambda_0$) is the wavenumber in free-space. The imaginary component is the extinction coefficient, $\kappa$, related to $\alpha$, and $\lambda_0$ as $\alpha=\frac{4\pi}{\lambda_0}{\kappa}$\cite{ReedGrahamT2004Sp:a}.

The simulated light propagation along the WG shows the electric field profile, Fig.\ref{fig:Figure 1}c, of the propagating mode. The mode interaction with SLG is increased by maximizing the overlap between SLG and field profile. The metal contacts for each SLG are placed 1$\mu$m away from the WG edge to avoid optical losses due to proximity of field profile and contacts. The EO response is simulated by varying $E_\text{F}$ between $0-0.8$eV and extracting the change in $n_{\text{eff}}$ as a function of $E_\text{F}$. We use $E_\text{F}\sim$0.2eV for ungated SLG, to account for the impurity doping of as prepared SLG\cite{Sorianello2018,Muench2019}. Simulations are performed at 300K for $\lambda$=1.55$\mu$m with $\tau$=440,22,11fs, corresponding to $\mu\sim$10,000, 500, 250cm$^2$/Vs at $E_\text{F}\sim$0.4eV. We then calculate the phase shift $\Delta\phi=k_0\Delta n_{\text{eff}}L$ and optical losses $I/I_0=e^{-\frac{4\pi}{k}L}$ induced by SLG for a given $L$\cite{ReedGrahamT2004Sp:a}. We relate $E_\text{F}$ to the applied voltage, $V$, by considering the sum of the potential applied across the overlapping SLG regions and the surface voltage due to the accumulated charges at each SLG electrode\cite{HuChenming2010Msdf,Sorianello2015}:
\begin{equation}
|V - V_\text{CNP}| = \frac{e}{C_\text{{eq}}}\frac{1}{\pi}\left(\frac{E_\text{F}}{\hbar v_F}\right)^2 + 2 \frac{|E_\text{F}|}{e}
\end{equation}

The potential across the overlapping SLG region is related to the total number of accumulated charges, $n_{\text{tot}}$, and the equivalent capacitance, $C_\text{eq}$ of the overlapping SLG region. $C_\text{eq}$ is the series combination of the quantum capacitance\cite{Luryi1988}, $C_Q$, of each SLG and the capacitance of the parallel-plate geometry, $C_\text{ox}$. The equivalent electrical circuit of the DSLG modulator is in Fig.\ref{fig:Figure 1}b. $C_Q=2e^2\sqrt{n_{\text{tot}}}/\hbar v_F\sqrt{\pi}$\cite{Xia2009,Das2009} and $C_\text{ox}=\epsilon_r\epsilon_0/d$, where $d$ is the thickness of the gate oxide separating the two SLG, $\epsilon_r$ is the relative permittivity of Al$_2$O$_3$, and $\epsilon_0$ is the permittivity of free space. The dielectric constant of Al$_2$O$_3$ is measured with a Woollam Ellipsometer M-2000 as $\epsilon_r\sim8$. To account for charged impurities at the SLG-Al$_2$O$_3$ interface and the impurities introduced during growth or device fabrication, we model the total charge density $n_{\text{tot}}$ as the sum of a carrier concentration from electrostatic doping $n(V_G)$, where V$_G$ is gate voltage, and an additional concentration from charged impurities $n_{\text{imp}}$\cite{Xia2009}. A charged-impurity density$\sim10^{12}\text{cm}^{-2}$\cite{Hwang2007}, leads to an increase in carrier concentration of SLG$\sim10^{11}\text{cm}^{-2}$\cite{Xia2009}.

Fig.\ref{fig:Figure 1}d plots the simulated EO response at 1550nm in terms of $\Delta n_{\text{eff}}$ and associated optical losses per $\mu$m with increasing $V_G$ and $E_\text{F}$. Optical losses decrease when $E_\text{F}>$0.2eV, corresponding to intraband transitions and the onset of Pauli blocking. For $E_\text{F}>$0.6eV, SLG enters the transparency regime where interband transitions are blocked such that optical losses are minimised and do not change as $E_\text{F}$ is further increased. $\Delta n_\text{eff}$ changes sign with increasing $V_G$, giving a positive or negative $\Delta\phi$ for the modulated signal. A bias voltage can be applied to the DSLG modulator to define the operating point on the EO response curve in Fig.\ref{fig:Figure 1}d. The amplitude of the driving voltage defines the operating range around the operating point. The ideal working point for pure PM is in the transparent region where $\Delta n_\text{eff}$ changes quasi-linearly, whilst optical losses remain constant. This also minimises power consumption because optical losses are at their lowest. Optical losses depend on $\tau$, as plotted in Fig.\ref{fig:Figure 1}d for $\tau=$440, 22, 11fs. Low scattering time $\tau$ is associated with high scattering rate $\Gamma$ via $\tau=\hbar/\Gamma$\cite{DasSarma2011}, leading to increased absorption via intraband transitions and reduced absorption via interband transitions. As $E_\text{F}$ approaches 0.4eV, optical losses are reduced for a lower $\tau$ because absorption via interband transitions is reduced. However, in transparency, increased intraband transitions lead to optical losses over 3 times greater for $\tau$=11fs, when compared to 440fs.
\begin{figure*}
\centerline{\includegraphics[width=\textwidth]{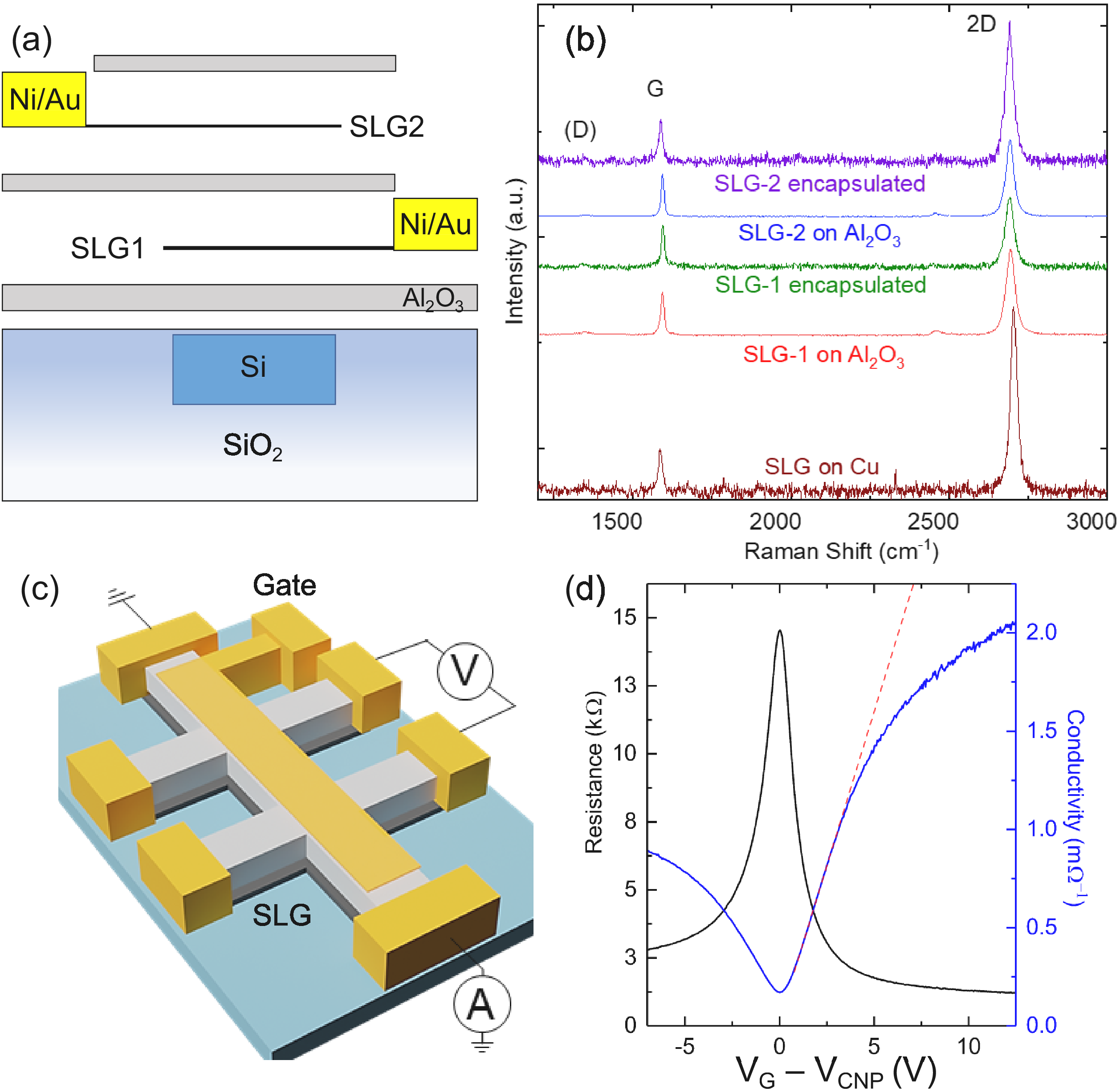}}
\caption{(a) DSLG modulator fabrication schematic: 10nm Al$_2$O$_3$ deposition on Si WG with 20nm Al$_2$O$_3$ between bottom and top SLG. 10nm Al$_2$O$_3$ is used to encapsulate both SLGs to protect them during subsequent processing steps, minimise contamination, and preserve $\mu$. The bottom encapsulation is used to maintain symmetry between the two SLGs, so that both are in the same environment. (b) Raman spectra at 514nm for the SLG closest to the WG (SLG1) and that farthest from the WG (SLG2), as-grown on Cu, after transfer, after device fabrication. The spectra are normalised to I(G), with Cu background PL removal\cite{Lagatsky2013}. (c) Schematic of SLG (grey) on SiO$_2$ (green) top-gated Hall bar device with Ni/Au (yellow) contacts. d) Measured R$_S$ (black) and calculated $\sigma_\text{d.c.}$ ($\sigma_\text{d.c.}=1/R_S$) (blue). Red dashed line is the $\sigma_\text{d.c.}$ linear fit for$V>$0, showing the transition from the linear to sub-linear regime for $V>$5V.}
\label{fig:Figure 2}
\end{figure*}

The speed of the DSLG phase modulator is defined by the cut-off frequency, $f_{3\text{dB}}$, at which the power of the modulated signal has decreased by half (3dB)\cite{Reed2010}. The dominant factor that limits $f_{3\text{dB}}$ is the product of the circuit resistance, $R$, and capacitance, $C$, known as the RC response\cite{HorowitzP1980Taoe}. We estimate this by electrical modelling, considering the different contributions to the total circuit impedance $Z_\text{T}(\omega)$, coming from each contact, $R_C$, ungated SLG sections, $R_\text{ungated}$, and gated SLG sections, $R_\text{gated}$. The equivalent circuit, Fig.\ref{fig:Figure 1}b, contains these components in series:
\begin{equation}
Z_\text{T}(\omega) = 2(R_\text{C} + R_\text{{ungated}}) + Z_\text{C}(\omega)
\end{equation}
Where $Z_\text{C}(\omega)$ is the impedance of the overlapping SLG regions. $Z_\text{C}(\omega)$ is given by $C_\text{eq}$ in series with $R_\text{{gated}}$ for each SLG electrode. The resistance ($R$) of SLG can be related to the sheet resistance ($R_S$) of SLG as $R=LR_S/w$ \cite{Bonaccorso2010} where $L$ and $w$ the length and width of SLG, respectively. By considering $\omega\to$0, $R_S$ can be related to the electrical conductivity $\sigma_\text{d.c.}$, as $R_S=1/\sigma_\text{d.c.}$\cite{KittelC1996Itss}. $R_S$ is calculated for different $E_\text{F}$ and $\tau$ from $\sigma_\text{d.c.}=ne\mu$\cite{Bonaccorso2010}. $R_S$ depends on $E_\text{F}$, therefore on the voltage applied across the DSLG modulator. From Ohm's law and $Z_\text{T}(\omega)$, we calculate the frequency dependent current, $I(\omega)$, flowing through the circuit at a nominal drive voltage, $V_\text{D}$, $I(\omega)=V_\text{D}/Z_\text{T}(\omega)$. The voltage drop across the DSLG modulator, $V_\text{M}$, is $V_\text{M}(\omega)=I(\omega)/j\omega C_\text{eq}$. When $I(\omega)$ starts to flow and $V_\text{M}(\omega)$ goes to 0, the modulator is no longer operational. Therefore, by substituting $I(\omega)$ into $V_\text{M}(\omega)$, we get:
\begin{equation}\label{eq:RC_response}
\frac{V_\text{M}(\omega)}{V_\text{D}}=\frac{1}{1+j\omega RC_\text{eq}}
\end{equation}
\begin{table*}[t]
    \centering
    \begin{tabular*}{\linewidth}{@{\extracolsep{\fill}}cccccccccccccc}
    \toprule % Top horizontal line
    Samples & SLG1 & Encapsulated SLG1 & SLG2 & Encapsulated SLG2 \\ % Column names row
    \midrule % In-table horizontal line
    Pos(G) (cm$^{-1}$) & 1592 $\pm$ 3 & 1596 $\pm$ 1 & 1596 $\pm$ 1 & 1585 $\pm$ 5 \\ % Content row 1
    FWHM(G) (cm$^{-1}$) & 14 $\pm$ 3 & 12 $\pm$ 2 & 11 $\pm$ 2 & 17 $\pm$ 2 \\ % Content row 2
    Pos(2D) (cm$^{-1}$) & 2692 $\pm$ 3 & 2691 $\pm$ 1 & 2694 $\pm$ 2 & 2689 $\pm$ 2 \\ % Content row 3
    FWHM(2D) (cm$^{-1}$) & 31 $\pm$ 1 & 30 $\pm$ 1 & 29 $\pm$ 3 & 30 $\pm$ 4 \\ % Content row 4
     A(2D)/A(G) (cm$^{-1}$) & 2.2 $\pm$ 0.4 & 5.6 $\pm$ 1.7 & 2.7 $\pm$ 0.1 & 6.3 $\pm$ 0.7 \\ % Content row 5
     I(2D)/I(G) (cm$^{-1}$) & 2.7 $\pm$ 0.7 & 2.4 $\pm$ 1 & 1.8 $\pm$ 0.1 & 3.5 $\pm$ 0.5 \\ % Content row 6
     I(D)/I(G) (cm$^{-1}$) & 0.07 $\pm$ 0.03 & 0.03 $\pm$ 0.04 & 0.05 $\pm$ 0.04 & 0.13 $\pm$ 0.13\\ % Content row 7
     E$_F$ (meV) & 190 $\pm$ 80 & 276 $\pm$ 158 & 292 $\pm$ 87 & 180 $\pm$ 130 \\ % Content row 8
     Doping type & p & p & p & p \\ % Content row 9
     n (x 10$^{12}$) (cm$^{-2}$) & 2.6 $\pm$ 2.0 & 8.5 $\pm$ 9.5 & 5.8 $\pm$ 3.2 & 3.9 $\pm$ 3.8 \\ % Content row 10
     Uniaxial strain ($\%$) & -0.20 $\pm$ 0.32 & -0.15 $\pm$ 0.18 & -0.07 $\pm$ 0.17 & 0.08 $\pm$ 0.07 \\ % Content row 11
     Biaxial strain ($\%$) & -0.08 $\pm$ 0.14 & -0.06 $\pm$ 0.07 & -0.02 $\pm$ 0.06 & 0.003 $\pm$ 0.02 \\ % Content row 12
     n$_D$ (x 10$^{10}$) (cm$^{-2}$) & 2.6 $\pm$ 0.4 & 1.6 $\pm$ 0.9 & 2.5 $\pm$ 0.6 & 4.2 $\pm$ 2.7 \\ % Content row 13
    \bottomrule % Bottom horizontal line
    \end{tabular*}
    \caption{Raman fit parameters and corresponding E$_F$, doping type, $n$, strain, n$_D$, and error bars.}
    \label{tab2:RamanFits}
\end{table*}

The simulated $f$ response of our DSLG modulators is in Fig.\ref{fig:Figure 1}e. $f_\text{3\text{dB}}$ increases with $\mu$ in Fig.\ref{fig:Figure 1}f due to a reduction in $R_\text{ungated}$ and $R_\text{gated}$ for SLG with higher $\mu$. Assuming a constant $\mu$, $f_\text{3\text{dB}}$ can be increased by reducing $C$ and $R$. However, there is a trade-off between minimising ungated SLG length, to reduce $R$, and minimising the gated SLG length, to reduce $C$. Even though $R_\text{ungated}>R_\text{gated}$, $R_\text{gated}$ is of the same order of magnitude as $R_C$. $f_\text{3\text{dB}}$ can be further increased by minimising the distance between contacts and WG, to minimise the impact from ungated regions. There is a trade-off between minimising the required $V$ to reach Pauli blocking and maximising $f_\text{3\text{dB}}$. To reduce $V$, $C_\text{ox}$ should be maximised by using a dielectric with the highest $\epsilon_r$ or reducing $d$. However, to increase $f_\text{3\text{dB}}$, $C_\text{ox}$ should be reduced by increasing $d$ and reducing the size of overlapping SLG region. We limit the size of overlapping SLG to the WG width and use 20nm Al$_2$O$_3$ to maximise $f_\text{3\text{dB}}$ and limit $V<$15V. To operate in the transparency regime, the dielectric should support the required $V$ to reach $E_\text{F}>$0.4eV without breakdown. Minimising the size of the overlapping SLG region, we reduce the probability of breakdown due to pinholes in the dielectric.

The DSLG modulators are then fabricated as for Fig.\ref{fig:Figure 2}a. We use the IMEC silicon-on-insulator (SOI) platform because of the low (2.3dB per grating) coupling losses\cite{Absil2015}. 10nm Al$_2$O$_3$ is deposited on SOI by atomic layer deposition (ALD, Cambridge Nanotech Savannah S100 G1) at 120$^o$C. After a 10min purge of N$_2$ for contaminants removal, we apply 238 consecutive cycles of 22ms pulses of deionized water and 17 ms pulses of trimethylaluminum precursors to reach the required 10nm thickness, as measured with a Woollam Spectroscopic Ellipsometer M-2000XI. Continuous SLG is grown on Cu by chemical vapor deposition (CVD). The Cu foil is first annealed at 1050$^o$C under 90\% H$_2$ and 10\% Ar at 760torr for 2h and cooled to $RT$. To grow SLG, the annealed Cu foil is heated to 1050$^o$C with 40sccm H$_2$ at 0.4Torr and annealed for 2h. Growth is initiated by introducing 5sccm CH$_4$ and the CH$_4$ flow is stopped to terminate growth after 30mins, and SLG/Cu is cooled to $RT$\cite{Li2009a}. SLG is then wet-transferred using polymethyl-methacrylate (PMMA) as a supporting layer and Cu etching in ammonium persulfate\cite{Bonaccorso2012}.
\begin{figure*}
\centerline{\includegraphics[width=\textwidth]{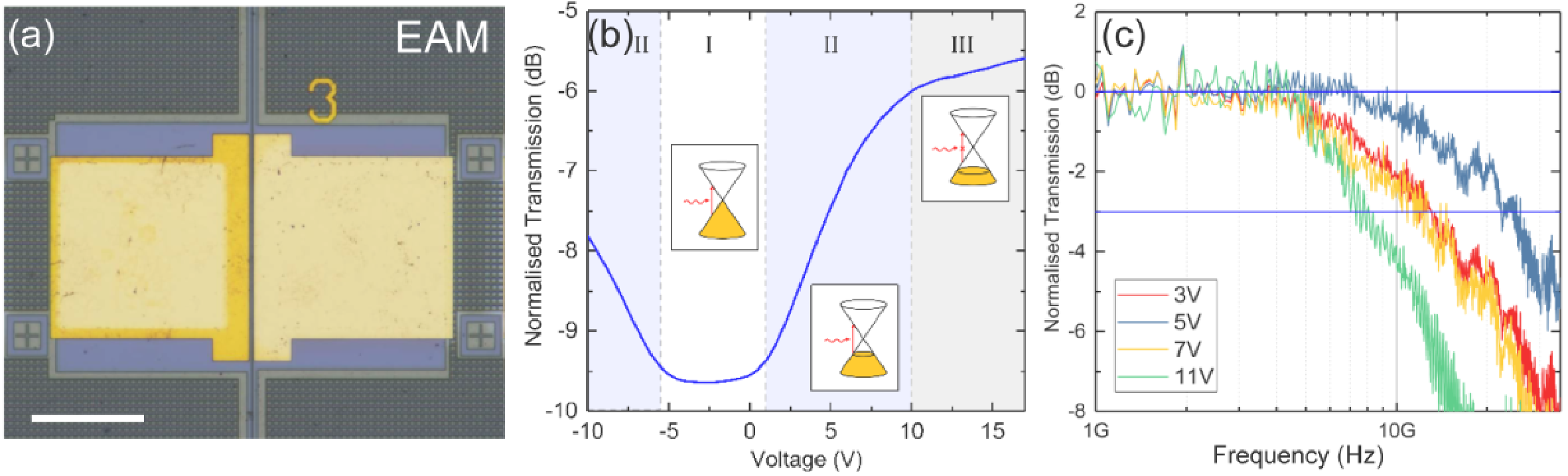}}
\caption{(a) Optical micrograph of EAM consisting of DSLG modulator on straight Si WGs. (b) EO response of a 75$\mu$m DSLG EAM showing the different regimes depending on $E_\text{F}$. At 1.55$\mu$m, transmission is lowest in region I (white), when $E_\text{F}<0.4$eV. It increases in region II (blue) due to the onset of Pauli blocking, when $E_\text{F}$ approaches 0.4eV, before transitioning to the minimum-loss regime in region III (grey)$>$10V, where $E_\text{F}>$0.4eV. The Si WG is 450x220nm$^2$. (c) EO frequency response for a 50$\mu$m DSLG EAM at different DC biases with a 1V peak-to-peak driving voltage. $f_\text{3\text{dB}}\sim$13GHz for 3V, 24GHz for 5V, 12GHz for 7V, and 8GHz for 11V.}
\label{fig:Figure 3}
\end{figure*}

As-grown and transferred SLG are characterised by Raman spectroscopy with a Renishaw InVia spectrometer equipped with 50x objective at 514.5nm. 6 spectra are collected from both as grown SLG on Cu and transferred SLG to estimate doping and defect density. The errors are calculated from the standard deviation across different spectra, the spectrometer resolution ($\sim$1cm$^{-1}$) and the uncertainty associated with the different methods to estimate the doping from full width at half maximum of G-peak, FWHM(G), intensity and area ratios of 2D and G peaks, I(2D)/I(G), A(2D)/A(G). Table \ref{tab2:RamanFits} summarises Raman peaks fit are derived E$_F$, doping type, charge carrier density $n$, strain, and defects density n$_D$. The Raman spectrum of as grown SLG is in Fig.\ref{fig:Figure 2}b, after Cu photo-luminescence removal\cite{Lagatsky2013}. The 2D peak is a single-Lorentzian with FWHM(2D)=27$\pm$2cm$^{-1}$, signature of SLG\cite{Ferrari2006,Ferrari2013}. Pos(G)=1591$\pm$4cm$^{-1}$ with FWHM(G)=16$\pm$2cm$^{-1}$. Pos(2D)=2712$\pm$9cm$^{-1}$, I(2D)/I(G)$\sim$4.5$\pm$0.7 and A(2D)/A(G)7.6$\pm$1.1. No D peak is observed, indicating negligible Raman active defects\cite{Ferrari2000, Cancado2011}. First, SLG1 is transferred on 10nm Al$_2$O$_3$ deposited on SOI. The representative Raman spectrum of transferred SLG1 before Al$_2$O$_3$ encapsulation is in Fig.\ref{fig:Figure 2}b. The 2D peak retains its single-Lorentzian line shape with FWHM(2D)=31$\pm$1cm$^{-1}$, Pos(G)=1592$\pm$3cm$^{-1}$, FWHM(G)=14$\pm$3cm$^{-1}$, Pos(2D)=2692$\pm$3cm$^{-1}$, I(2D)/I(G)=2.7$\pm$0.7, and A(2D)/A(G)=2.2$\pm$0.4 indicating a p-doping with E$_F$=190$\pm$80meV\cite{Das2008,Basko2009}. I(D)/I(G)=0.07$\pm$0.03 corresponds to a defect density n$_D\sim2.6\pm$0.4x10$^{10}$\cite{Bruna2014} for excitation energy of 2.41eV. SLG1 is then patterned by electron beam lithography (EBL) using a Raith EBPG5200, followed by a 60s O$_2$ plasma at 10W using a Vision 320 reactive ion etcher (RIE). Contacts are fabricated using a double-layer resist mask of Methyl methacrylate and PMMA\cite{Shaygan2017}, followed by 15/50nm Ni/Au deposited by sputter coating (Precision Atomics Metallifier Sputter Coater) and thermal evaporation (M-Braun PROvap PVD system). A 1nm seed-layer of Al is then thermally evaporated, before 20nm of Al$_2$O$_3$ is deposited by ALD at 120$^o$C on SLG1. After Al$_2$O$_3$ encapsulation, the 2D peak in SLG1 retains its single-Lorentzian line shape with FWHM(2D)=30$\pm$1cm$^{-1}$, Pos(G)=1596$\pm$1cm$^{-1}$, FWHM(G)=12$\pm$2cm$^{-1}$, Pos(2D)=2691$\pm$1cm$^{-1}$, I(2D)/I(G)=2.4$\pm$1, and A(2D)/A(G)=5.6$\pm$1.7 indicating a p-doping with E$_F$=276$\pm$158meV\cite{Das2008,Basko2009}. I(D)/I(G)=0.03$\pm$0.04 corresponds to n$_D$=1.6$\pm$0.9x10$^{10}$\cite{Bruna2014} for 2.41eV excitation. SLG2 is transferred using the same process as SLG1, and characterized by Raman spectroscopy (Fig.\ref{fig:Figure 2}b). The 2D peak retains its single-Lorentzian line shape with FWHM(2D)=29$\pm$3cm$^{-1}$. Pos(G)=1596$\pm$1cm$^{-1}$, FWHM(G)=11$\pm$2cm$^{-1}$, Pos(2D)=2694$\pm$2cm$^{-1}$, I(2D)/I(G)=1.8$\pm$0.1, and A(2D)/A(G)=2.7$\pm$0.1 indicating a p-doping with E$_F$=292$\pm$87meV\cite{Das2008,Basko2009}. I(D)/I(G)=0.05$\pm$0.04 corresponds to n$_D$=2.5$\pm$0.6x10$^{10}$\cite{Bruna2014} for 2.41eV. SLG2 is then patterned by using O$_2$ plasma after EBL and contacts are fabricated using a double-layer resist mask for EBL as SLG1, and subsequent Ni/Au (15/50nm) deposition. Finally, 10nm Al$_2$O$_3$ is deposited on SLG2 after a 1nm Al seed-layer is thermally evaporated. After Al$_2$O$_3$ encapsulation, the 2D peak in SLG2 retains its single-Lorentzian line shape with FWHM(2D)=30$\pm$4cm$^{-1}$, Pos(G)=1585$\pm$5cm$^{-1}$, FWHM(G)=17$\pm$2cm$^{-1}$, Pos(2D)=2689$\pm$2cm$^{-1}$, I(2D)/I(G)=3.5$\pm$0.5, and A(2D)/A(G)=6.3$\pm$0.7 indicating a p-doping with E$_F$=180$\pm$130meV\cite{Das2008,Basko2009}. I(D)/I(G)=0.13$\pm$0.13 gives n$_D$=4.2$\pm$2.7x10$^{10}$\cite{Bruna2014} for 2.41eV. SLG1 and SLG2 show different doping and defect density even though they are transferred from the same SLG/Cu because SLG1 is subject to more fabrication steps than SLG2. Strain is estimated from Pos(G)\cite{Mohiuddin2009, Yoon2011}. Biaxial strain can be differentiated from uniaxial by the absence of G-peak splitting with increasing strain, however at low ($\leq$0.5\%) strain the splitting cannot be resolved. For uniaxial (biaxial) strain, Pos(G) depends on both E$_F$ and strain\cite{Das2008,Mohiuddin2009}, to get the contribution of strain only, we first derive E$_F$ from A(2D)/A(G), I(2D)/I(G) and FWHM(G), which are independent of strain\cite{Das2008,Basko2009,Pisana2007}, and then calculate Pos(G) corresponding to this E$_F$. The strain is then retrieved from the difference between the experimental and calculated Pos(G), Table \ref{tab2:RamanFits}.

A 4-point-probe measurement using top-gated Hall bar structures (Fig.\ref{fig:Figure 2}c) is performed to derive SLG resistance and conductivity. Fig.\ref{fig:Figure 2}d plots the measured voltage-dependent resistance and the calculated $\sigma_\text{d.c.}$ after normalising the conductance to the channel geometry. We observe the expected\cite{Novoselov2004} peak in resistance, which corresponds to the SLG Dirac point. $\mu$ is estimated from the measured conductivity as $\sigma_\text{d.c.}=ne\mu$\cite{KittelC1996Itss, Novoselov2004}, where the linear region of $\sigma_\text{d.c.}$ in Fig.\ref{fig:Figure 2}d corresponds to a constant $\mu$. The charge density $n$ in terms of $C_\text{ox}=\epsilon\epsilon_\text{0}/d$ can be written as $n=C_\text{ox}(V_\text{G}-V_\text{CNP})/e$\cite{Novoselov2004}, hence the conductivity becomes $\sigma_\text{d.c.}=\epsilon\epsilon_\text{0}\mu(V_\text{G}-V_\text{CNP})/d$. Using measured dielectric constant and thickness of $Al_2O_3$ $\epsilon_\text{Al2O3}=8$ and d=20nm, the measured $\sigma_\text{d.c.}$ can be fitted to estimate $\mu\sim1000cm^2/Vs$. The linear fit to SLG conductivity as a function of V$_G$ is in Fig.\ref{fig:Figure 2}d.
\begin{figure*}
\centerline{\includegraphics[width=\textwidth]{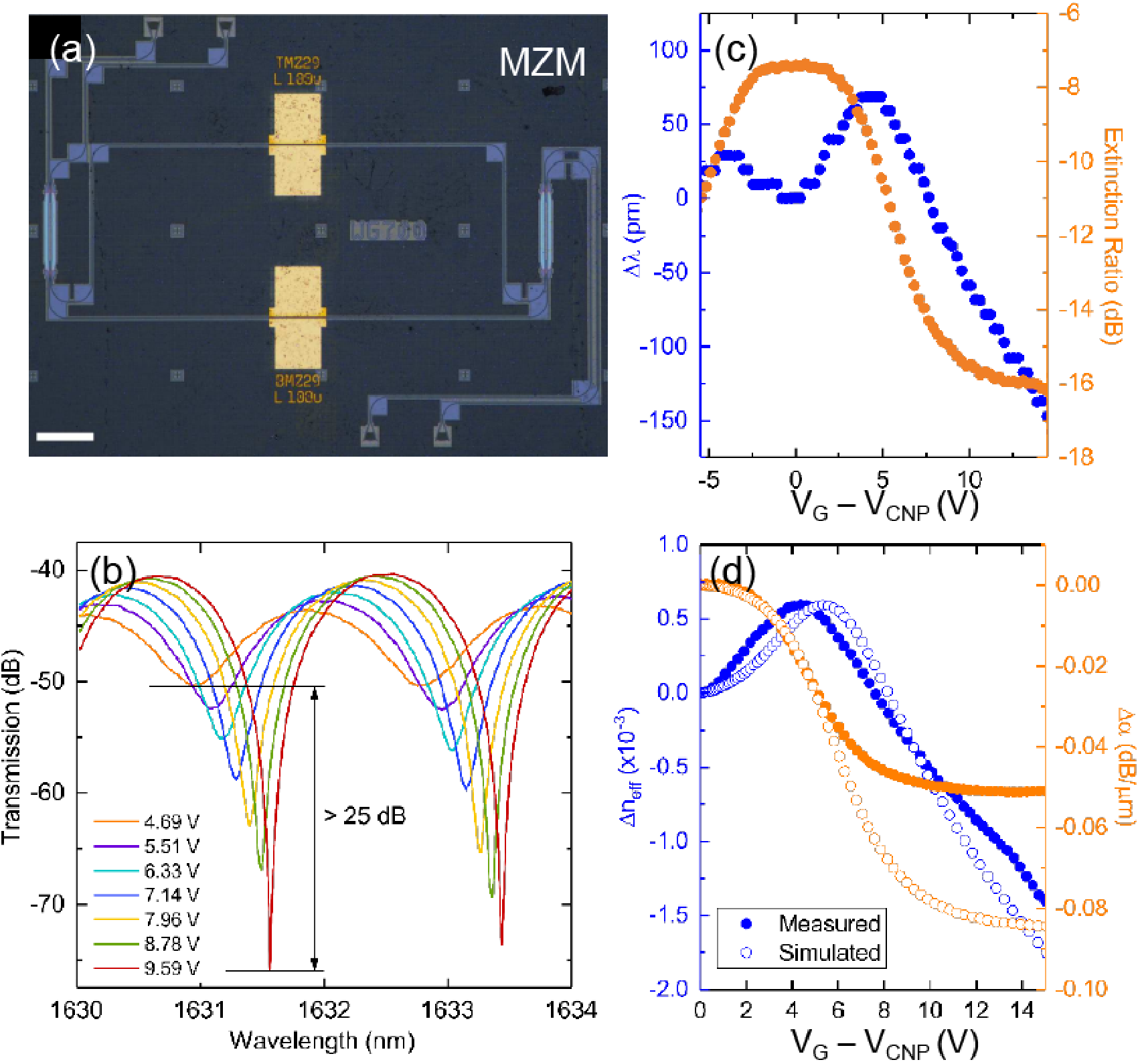}}
\caption{(a) Optical micrograph of MZM consisting of a DSLG modulator on each arm of a Si MZI. Scale bars 100 $\mu$m. (b) Voltage-dependent transmission of a MZM containing one 450$\mu m$ DSLG modulator on each arm, one biased at 10V and the other swept from 4V (orange) to 10V (red). (c) Voltage-dependent shift of interference fringe position (blue) and ER (orange) of an MZM with a$\sim$100$\mu$m DSLG modulator on each arm, with one modulator biased at 10V. The MZI is unbalanced, with 2 input and output ports, and 620x220 nm$^2$ Si WGs. (d) Comparison of measured (solid circles) and simulated (open circles) $\Delta n_\text{eff}$ (blue) and $\Delta\alpha$ (orange) for a 450$\mu$m DSLG MZM. Simulation performed at 1.55$\mu$m with $\tau$=14fs with same structure as the measured device, 550x220 nm$^2$ Si WG, overlapping SLG region$\sim$550nm, ungated SLG region$\sim$1$\mu$m ($E_\text{F}$=0.2eV), 20nm Al$_2$O$_3$ with $\epsilon_\text{Al2O3}$=8.}
\label{fig:Figure 4}
\end{figure*}

The EO response of DSLG EAMs and ERMs are then measured using angled single-mode optical fibres to couple light into the photonic circuits via grating couplers. A representative EAM optical microscopy image in Fig.\ref{fig:Figure 3}a. The position of the fibres and the polarisation of the source laser (Agilent 8164B Lightwave Measurement System) are adjusted to minimise coupling losses and maximise the power coupled into the confined optical mode. The steady-state response is measured by applying a DC voltage across both SLGs and measuring the optical power at the output, $P_\text{out}$. The transmitted power, $P_\text{t}=10\log(P_\text{out}/P_\text{in})$, is expressed in dB. Fig.\ref{fig:Figure 3}b is the optical transmission of a DSLG EAM comprising a 75$\mu$m modulator on a straight WG. The EAM is biased between -10 and 17V at 1.55$\mu$m with $P_\text{in}$=1mW. To extract IL, $P_\text{t}$ needs to be normalised to account for the additional propagation and coupling losses introduced from processing. The increase in power loss compared to the loss before processing is due to the deposited Al$_2$O$_3$ on the grating couplers, residues from SLG transfer and device fabrication. The additional losses can be subtracted by measuring the transmission through a similar WG with same dimensions and grating couplers, that has undergone the same processing steps as the DSLG modulator. The lowest V$_G$ dependent transmission in EAM occurs in region I in Fig.\ref{fig:Figure 3}b, between -5 and 0V, when $E_\text{F}$ is less than the half of the photon energy, $\hbar\omega_\text{1550}/2\sim0.4$eV. Therefore, interband transitions in SLG are allowed in region I. The transmitted power is minimum in this voltage range around $V_\text{CNP}$ where E$_F$ is closer to Dirac point\cite{Li2008}. For intrinsic SLG, where $V_\text{CNP}$ coincides with $V_\text{G}$=0, the transmission curve would be centred at 0V. In Fig.\ref{fig:Figure 3}b the $V_\text{CNP}$ is$\sim$-2.5V. This corresponds to $E_\text{F}\sim$274meV, which represents the average E$_F$ of both SLGs in the DSLG EAM. This is also consistent with the average SLGs $E_\text{F}=275\pm170$meV, estimated from Raman spectroscopy of SLG1 and SLG2 after $Al_2O_3$ encapsulation, Table \ref{tab2:RamanFits}, Fig.\ref{fig:Figure 2}b. As $V_\text{G}$ increases in region II, $E_\text{F}$ approaches $\hbar\omega_\text{1550}/2$, and transmission increases due to Pauli blocking of inter-band transitions. For $V_\text{G}>$10V, the transmission plateaus when $E_\text{F}>\hbar\omega_\text{1550}/2$ and SLG enters the transparent regime.

In transparency, IL$\sim$5.6dB for the DSLG phase modulator, corresponding to loss$\sim$746dB/cm when normalised by the modulator length. This is higher than state-of-the-art Si depletion ($\sim$22dB/cm\cite{Li2018}), III-V ($\sim$19dB/cm\cite{Han2017}), LN ($\sim$0.25dB/cm\cite{Wang2018a}), and SLG ($\sim$236dB/cm\cite{Sorianello2018}) MZMs, Table \ref{tab1:Modulators}. We attribute the additional optical losses to scattering from resist residues and defects generated in SLG during fabrication, degrading $\tau$. The simulated optical loss of the same device structure is$\sim$93dB/cm for $\tau$=440fs. Therefore, IL can be further reduced by improving SLG processing, increasing $\mu$, hence reducing short-range scattering, and by developing a selective planarization process which isolates passive sections of the WGs. The size of overlapping SLG regions can be increased to further reduce IL, as the ungated SLG sections are not in transparency, hence contributing optical losses. However, any increase in the overlapping SLG region, will increase $C_\text{eq}$, therefore will reduce $f_\text{3\text{dB}}$. Increasing the overlapping SLG region to$\sim$1$\mu$m, would decrease optical loss to$<$10dB/cm for $\tau$=440fs, leading to IL$<$1dB for$\sim$3mm devices, matching IL of LN\cite{Wang2008} and III-V\cite{Han2017} MZMs, Table \ref{tab1:Modulators}.

The EO BW, or speed, is then measured by applying a sinusoidal voltage to the DSLG modulators, in either EAM or ERM configuration. The voltage is provided by a signal generator (Agilent E8257D PSG) combined with a DC voltage via a bias tee. The optical output from the DSLG modulator is then amplified with an Er doped fibre amplifier (EDFA, Keyopsys CEFA-C-HG) followed by a 1nm narrow-band optical filter, before going into a InGaAs photodetector (PD) with a BW$>$40GHz (Newport 1014). The narrow-band filter is used to remove the noise resulting from the spontaneous emission from the EDFA\cite{Mears1987}, and to ensure that the PD input power is below the safe input power=5mW given by the specifications of the Newport PD\cite{NewportPDSpecs}. The modulated output signal is recorded on an electrical spectrum analyser (ESA, Agilent PSX N9030A). By monitoring the amplitude of the modulated signal with increasing $f$, we get $f_\text{3\text{dB}}$. The setup is calibrated by repeating the measurements with the same configuration, but with a Thorlabs LN05S-FC AM with a 3dB cut-off$\sim$40GHz. A final normalisation is then done for the $f$ response of the Thorlabs LN05S-FC modulator, taken from the supplied data sheet\cite{Uehara1978}. Fig.\ref{fig:Figure 3}c is the $f$ response of a 50$\mu$m DSLG EAM for different DC biases. $f_\text{3\text{dB}}$ increases from 13 to 25GHz between 3 and 5V, then decreases to 12GHz for 7V, and 8GHz for 11V. We attribute the decrease in $f_\text{3\text{dB}}$ above 5V to a reduction in $\mu$ due to increased short-range scattering of charge carriers as $E_\text{F}$ increases. For our RC limited devices, we expect $f_\text{3\text{dB}}$ to increase with $V$, because $R_S$ reduces with increasing $V$\cite{Novoselov2004}, up to an optimum point after which the increase of $C_q$ becomes predominant. The optimal bias point for operating our device at 25 GHz BW is$\sim5V$, whilst operating it at CMOS-compatible voltages ($<$2V) allows 13GHz, Fig.\ref{fig:Figure 3}c. We assign the $V_G$-dependent slow-down in Fig.\ref{fig:Figure 3}c to a decrease in $\mu$ above 5V due to increased short-range scattering as V$_G$ increases. This would lead to a $f_\text{3\text{dB}}$ slow-down of the same order of magnitude as that measured between 5 and 11V, where $f_\text{3\text{dB}}$ drops from 25 to 8GHz. This contrasts the increase from 13 to 25GHz between 3 and 5V, where we are still in the linear region of $\sigma_\text{d.c.}$, and benefit from decreasing $R_S$. The transition to sub-linear behaviour can be pushed to higher V$_G$ by decreasing the sources of short-range scattering in SLG, from SLG processing improvements. Thus, $f_\text{3\text{dB}}$ can be increased by improving SLG growth and transfer, to limit $\mu$ degradation during fabrication.

The simultaneous phase change that accompanies the change of amplitude cannot be extracted from an electro-absorption configuration, because the transmission of a straight WG is independent of optical signal phase\cite{ReedGrahamT2004Sp:a}. Instead, it is measured using an electro-refractive configuration, with a Mach-Zehnder interferometer (MZI)\cite{ReedGrahamT2004Sp:a,MadsenChristiK1999Ofda}. The optical microscopy image of a representative MZI is in Fig.\ref{fig:Figure 4}a. Here, the optical signal is split into two arms. Depending on the phase difference between these, $\Delta\phi$, the propagating waves will interfere when recombined. If the propagating waves are in phase, transmission will not depend on $\lambda$\cite{ReedGrahamT2004Sp:a}. If they are not in phase, an interference pattern will appear\cite{ReedGrahamT2004Sp:a}. This is characterised by the free spectral range ($FSR=\frac{\lambda_\text{res}^2 n_g}{\Delta L}$\cite{MadsenChristiK1999Ofda}), defined as the wavelength difference between each transmission minima, where $\lambda_\text{res}$ is the fringe position, $n_g$ is the group index, and $\Delta L$ is the length difference between the arms of the MZI. $\lambda_\text{res}$ depends on $\Delta\phi$, which can result from a modulator that induces $\Delta n_\text{eff}$, or when the MZI arms are different lengths, known as an unbalanced MZI\cite{ReedGrahamT2008Sp:t}. A MZM uses an ERM on one or both MZI arms to control $\Delta\phi$. $\Delta\phi$ can then be directly measured by the shift of the output interference pattern\cite{ReedGrahamT2004Sp:a}. Placing an ERM on each arm enables the phase to be controlled independently on each arm to reach the required $\Delta\phi$. Fig.\ref{fig:Figure 4}b shows the V-dependent transmission of a MZM with a 450$\mu$m DSLG modulator on each arm of an unbalanced MZI. One device is biased at 10V, so that it is in the transparency regime, and the other is swept from 4 to 10V. By measuring the fringe shift, $\Delta\lambda$, for different V$_G$, we determine $\Delta\phi$. The measured $\Delta\lambda$ is normalised by the FSR, which corresponds to a phase difference of 2$\pi$\cite{ReedGrahamT2004Sp:a}, giving $\Delta\phi$ in units of $\pi$: $\Delta\phi[\pi]=\Delta\lambda/(\text{FSR}/2)$\cite{Mohsin2015}. $\Delta\phi$ is related to the V-induced change in the real component of $n_\text{eff}$ along L, by $\Delta\phi=k_0\Delta n_\text{eff}L$\cite{ReedGrahamT2004Sp:a}.

The extinction ratio (ER)=$10\log(P_\text{t,max}/P_\text{t,min})$\cite{ReedGrahamT2008Sp:t} is affected by the difference in absorption between the MZI arms, $\Delta\alpha$. If the propagating wave in one arm is absorbed, there is no interference at the output, because there will only be one propagating wave remaining\cite{ReedGrahamT2004Sp:a}. For losses that do not result in complete absorption, ER will increase when $\Delta\alpha$ is minimised, and decrease when $\Delta\alpha$ is maximised. The MZM ER can be related to $\Delta\alpha$ by considering the transmission through the MZM as the sum of the electric fields propagating down each MZI arm: $E\propto e^{-\alpha}\sin({\omega t-\phi})$\cite{ReedGrahamT2004Sp:a,MadsenChristiK1999Ofda}. The ER is given by the ratio of the maximum and minimum transmission through the MZM, which occurs when $\Delta\phi=0$ and $\pi$\cite{ReedGrahamT2004Sp:a}:
\begin{equation}
ER=\frac{1+e^{2\Delta\alpha}+2e^{\Delta\alpha}}{1+e^{2\Delta\alpha}-2e^{\Delta\alpha}}
\end{equation}
Fig.\ref{fig:Figure 4}b shows that, as $V_G$ increases, and the SLG on the active arm becomes transparent, ER increases$>$25dB because $\Delta\alpha$ is reduced. The effect of $\Delta\phi$ and $\Delta\alpha$ due to the active arm of the MZM is seen by the simultaneous change in position and ER of the interference fringes with $V_G$. Fig.\ref{fig:Figure 4}c plots the change in position of interference fringe and ER as a function of $V_G$. The MZM has the same behaviour as the EAM in Fig.\ref{fig:Figure 3}a. ER is minimised near the Dirac point, because the absorption of the SLG on the active arm is highest, whilst the device on the other arm is transparent. ER then increases with increasing $V_G$, as absorption by the active arm is reduced, until flattening$>$10V, when the SLGs on both arms are transparent. This shows that the transparency regime is ideal for pure PM because we have a quasi-linear change in phase, whilst losses remain constant. A similar V-dependent change in fringe position and ER is observed on either side of the Dirac point, where negative V$_G$ give a weaker effect than positive ones. The similarity around the Dirac point is due to the SLG ambipolarity\cite{Novoselov2004}, and the asymmetry can be due to different scattering rates of electrons ($e$) and holes ($h$), resulting from an uneven distribution of positively or negatively charged impurities\cite{Hwang2007,Tan2007,Chen2008,Anicic2013}. From $\Delta\lambda$ and $\Delta$ER we extract $\Delta n_\text{eff}$ and $\Delta\alpha$, Fig.\ref{fig:Figure 4}d. The measured and simulated $\Delta n_\text{eff}$ and $\Delta\alpha$ are in Fig.\ref{fig:Figure 4}d. We attribute the differences in measured and simulated behaviour to asymmetries between the two SLG as each SLG undergoes different amounts of processing, since SLG1 is subject to more processing than SLG2. The difference between measured and simulated $\Delta\alpha$ in transparency is a result of increased propagation losses outside the DSLG modulator, due to residues remaining on the WG from SLG processing. In transparency, the measured $\Delta\phi$ gives a $V_\pi L\sim$0.3V$\cdot$cm, matching that of state-of-the-art SLG PMs\cite{Sorianello2018}. However, unlike Ref.\citenum{Sorianello2018}, our devices have pure PM with negligible change in optical losses, an essential property for IQ modulation\cite{KumarShiva2014FOC}. Our DSLG MZMs have a $V_\pi L\sim$ on par with the lowest reported plasmonic LN MZMs\cite{Thomaschewski2022},$\sim$5 times better than the lowest reported LN MZMs\cite{Wang2018}, and$\sim$2.5 times better than the lowest reported thin film LN MZMs\cite{Li2023} and Si MZMs\cite{Liu2020}, Table \ref{tab1:Modulators}. Due to the high ($\sim$746dB/cm) optical loss, our DSLG phase modulator has FOM$_\text{PM}>$200VdB, greater than the lowest reported Si ($\sim$38VdB\cite{Li2018}), LN ($\sim$0.35VdB\cite{Wang2018a}), and III-V ($\sim$1VdB\cite{Han2017}), Table \ref{tab1:Modulators}. However, if the optical losses of SLG in transparency are reduced$<$10dB/cm by increasing the overlapping SLG region and increasing $\tau$ to$>$300fs, corresponding to $\mu>$6,000cm$^2$V$^{-1}$s$^{-1}$, our low $V_\pi L$ would enable FOM$_\text{PM}\sim$3VdB. This is lower than both Si and LN MZMs, with$\sim$3mm devices instead$\sim$2cm. Even though III-V MZMs have the lowest FOM$_\text{PM}\sim$1VdB\cite{Han2017}, their BW is unsuitable for Tb/s data transmission because it is limited to the MHz range\cite{Han2017}. Shrinking device dimensions by one order of magnitude results in denser circuits that benefit from reduced overall power consumption by minimising the interconnects lengths.
\section{conclusions}
We reported DSLG MZMs showing pure PM in the transparency regime for $E_\text{F}>$0.4eV, with V$_\pi$L$\sim$0.3Vcm. We reached the transparency regime by device design and process optimisation, ensuring the dielectric can withstand the required 10V to reach $E_\text{F}>0.4$eV without breakdown. Our low $V_\pi L$=0.3Vcm means we are able to overcome the loss limitations of Si MZMs, deliver increased circuit densities compared to LN, and match the performance of III-V (InGaAsP) MZMs, without expensive fabrication requirements. Reaching transparency is critical for graphene-based communications and metrology platforms that use complex modulation formats to maximise the density of transmitted information.
\section{acknowledgments}
We acknowledge funding from EU Graphene Flagship, ERC Grants Hetero2D, GIPT, EU Grants GRAP-X, CHARM, EPSRC Grants EP/K01711X/1, EP/K017144/1, EP/N010345/1, EP/L016087/1, EP/V000055/1, EP/X015742/1, Nokia Bell Labs, and MUR - Italian Minister of University and Research under the "Research projects of relevant national interest-PRIN 2020-Project No.2020JLZ52N "Light-matter interactions and the collective behavior of quantum 2D materials (q-LIMA)".

\end{document}